*Article*

# Autism Children's App using PECS

**Nareena Soomro[1, *] and Safeeullah Soomro[2]**

[1]Department of Computing Indus University, Karachi, Pakistan
nareena@indus.edu.pk
[2]Collage of Computer Studies, AMA International University, Kingdom of Bahrain
s.soomro@amaiu.edu.bh
*Correspondence: nainee_soom@yahoo.com

**Abstract:** Since autistic children suffers from learning disabilities and communication barriers, this research aim to design, develop and evaluate an Android based mobile application (app) providing better learning environment with inclusion of graphical representation in a cost effective manner. This research evaluate various supporting technologies and finds Picture Exchange Communication System (PECS) to be better choice for integrating with the app. Evaluation results reveal that the inclusion of PECS helped the children suffering from Autistic Spectrum Disorder (ASD) to better communicate with others. The study included autistic children who do not speak, who are unintelligible and who are minimally effective communicators with their present communication system. The evolution results showed encouraging impacts of the Autism App in supporting autistic children to adapt to normal life and improve the standard of their life.

## 1. Introduction

This research involved design, development and evaluation of an android based app using Picture Exchange Communication System (PECS) for the use of Autistic Spectrum Disorder (ADS) children. Autism is considered as a mental disorder. An autistic child do not tend to speak although the child is competent to spoke. The kid seems to be normal to a stranger, however, only after spending a good amount of time the problem(s) can be noticed. Kids, affected by autism, sometimes get panicked to have conversation and feel shy to speak. This is thus very important to provide the autistic kids with proper training, treatment, attention and care to help them cure. The term "autism" was derived from Greek word "autos" which means "self". Leo Kanner, doctor at Johns Hopkins University, initiated with term in 1940s to describe kid's behaviour which was socially and expressively inhibited [1]. Several investigators and psychotherapists are of the opinion that autism is linked disorder. Early treatment for autism included the use of electronic tremor, and manners change methods, which frequently depend on penalty and aching to change manners. Autism Spectrum Disorder children's and other evolving disabilities suffers from diverse communication difficulties [2]. The dearth variety from shortage of communication skills leads to difficulty in expending symbols and words to interact with another [3]. As part of this research, an android based mobile application (app) [4-6] for autistics children was developed which is aimed to function as an "assistant" to teachers, mentors, carers and doctors to help monitor and treat the autistic kids. The Autism app has used those technologies which is has already been being used for autistic children in hospitals and academia and considered to be useful for developing skills and teaching, such as Picture Exchange Communication System (PECS), certain phases etc. The app also provided the users with progress chart and user function for inclusion of pictures. PECS was established by MS. Lori Frost, Certificate of Clinical Competence/Speech Language Pathologists (CCC/SLP) and in 1984 Dr. Andrew Bondy used it on Delaware Autistic Platform [7-11]. It was used to teach kids with autism abstained, capability, efficient message system. PECS arise with conversation of uncomplicated imaged but rapidly build sentence construction. The use of PECS and indication semantic to teach autism children to interconnect, was highly opposed by some people at that time because they were of the opinion that these approaches were to hinder the growth of verbal language. However, based on the results of many research and studies, PECS truly helped people grow spoken language, reduce tantrums & strange manners and improved socializing. The Autism app followed certain phases along with progress which are considered to be supportive for teachers as well as for parents. These phases are as follows:

**How to Communicate (Phase I):** This phase emphasized on difficulties with common association that exists in autistic kids. Large number of children having autism experienced trouble with shared communication [12]. Although impairments in communication within inhabitants differ greatly, pupils study to conversation using particular images for actions.

**Discrimination between symbols (Phase II):** Students learned to make their picks from dissimilar objects. For instance, students were asked what is their favourite food and to identify apple and orange from two different pictorial cards. Through emblematic composition, the autistic kids were capable to composite actual life story with imaginary component, with knowledge that story was not real [13].

**Sentence Structure (Phase III):** Learners were trained to make easy sentences pursued by images of different things being appealed.

**Answering (Phase IV):** PECS was used to response the query, "What do you want?"

Autism App was based on HCI & Usability [14-15] convention and created extremely user friendly environment, interactive functions and attractive graphics and voice at the background behind every picture which might help the kid to speak such as alphabet. PECS is very simple as well as clear and contains Object represented as pictures in the cards that were used to express desire of autistic spectrum disorders (ASD) fatalities. The system was found to be extremely efficient in training essentials of language and assist autistic children to learn to communicate. In this phase, autism spectrum disorder person showed either typical practices in no less than one of the three classes: connection, correspondence, conduct and was normally analysed before the age of three [16]

## 1.1. Objectives

1. To conduct an expert survey to identify characteristics of the Autism children
2. To design and develop Android based mobile application with User Interface (UI) [17-19], integrated with the PECS, of Autism for children.

## 2. Autism App System

Authors of this paper conducted a survey study at different institutes, interviewing the trainers, teachers and doctors, distributing questionnaires. The information thus collected was used develop the app. Analysing the interview and survey results, the authors evaluated various development options and selected PECS (Picture Exchange Communication System) to be integrated in the app. PECS, having phases, uses cards to learn alphabets, words and make sentences using them and then move to motion, emotion and then sentence making structure. Audio cues [20-21] are associated with all the cards containing pictures or images. When the kids swipe the card, the cue associated with that card is played to help the kids to listen and learn the word/image of the card. The process learning process using PECS is more efficient when supervised by the parents and teachers (speech therapists).

None of the existing applications, such as [22] and [23], that were evaluated provide instant messaging functionality which has been integrated in our autism App. It provides the better user interfaces compared to other existing solutions. It is very simple and easy to navigate and also easier to visually analyse as it provides with big pictures on a white background with black text. In most of the existing apps, it is harder to navigate and the pictures are far too small, with words being the primary screen element, which doesn't fit the purpose of the application as the people using the application are more likely to recognize the picture representation. The goals of our Autism App include allowing the background of the application to be changed to make it easier to read, allowing the user to press on the card in order to read aloud the English representation of the card and a picture prediction function that will speed up the selection of symbols.

### 3. System Architecture

Upon start-up, the application promotes the Login/Registration screen so that the first time users can register and returning users can login. Upon successful login, the app provides with access to the main menu, as shown in Figure 1, which includes: (Single word learning) Pictures can be tapped to listen the words, (PECS Book) system have design communication book that kids learn to create a sentence, (Differentiate) user can drag and match the object getting stars in a progress and (Question/Answer) user can select the option in order to answer the question.

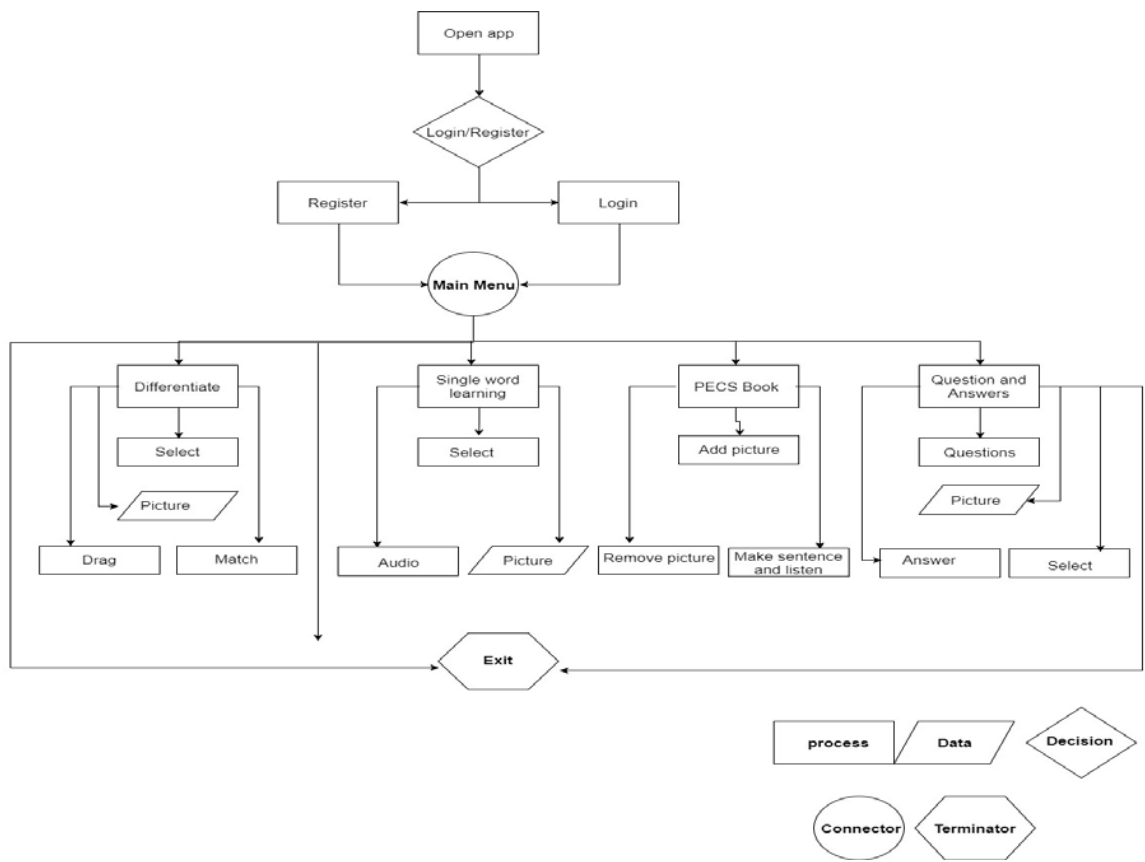

**Figure 1.** System Architecture

### 4. Proposed System

One of the options of the Autism App, as shown in Figure 2, is the Learning Single Word with subcategories such as Animals, Food, Colours, Shapes, Fruits, Emotions, Motions and Vegetables. Each subcategory comprises of reverent words with audio cues and pictures to learn by reading, visualising and listening. Using PECS Book functionality, user can select and remove cards to create & listen, make sentences and select suitable option to go to the next activity. Users can select and remove the suitable cards. Question/Answer requires that user has to select the right option/answer. The users are provided with feedback if they fail to select the right one. Scores are awarded in the form of starts upon selection of the right options/answers. The "Differentiate" activity provides the users to option to select similar or dissimilar ones from an array of objects. User can drag and match the right options/answers and get stars awarded.

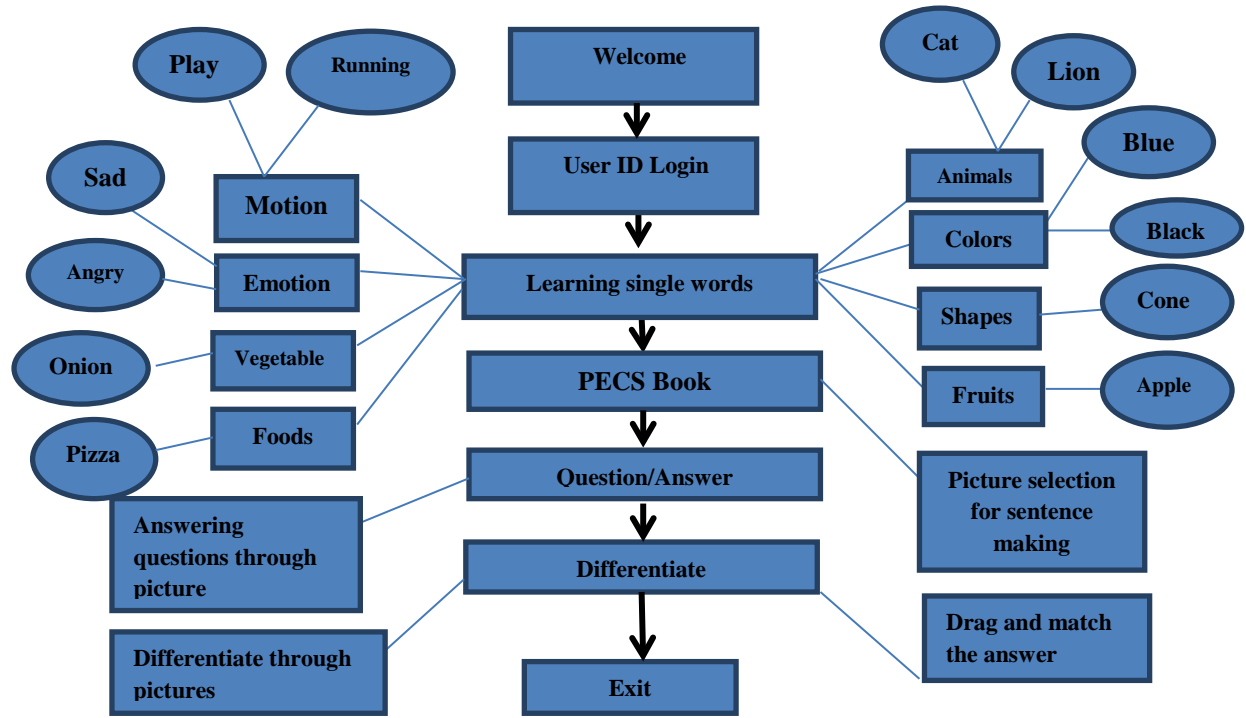

**Figure 2**. Proposed system of Activity

## 5. Design and Development

**Register:** The System, as shown in Figure 3a, provides the users the option to register for an account for enrolment. Basic demographic information, creation of username and password etc. are required to get registered using the system. Upon successful validation and verification process of the entered data, the accounts are created.

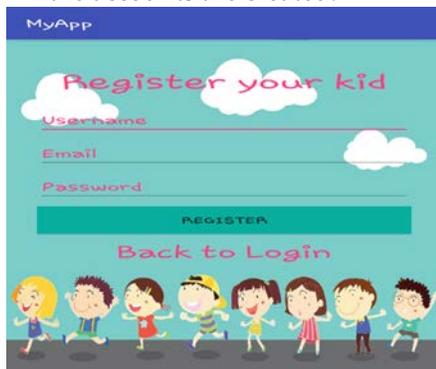

**Figure 3a.** Register

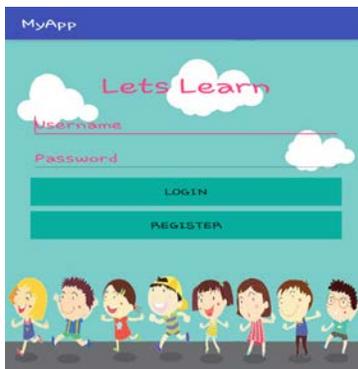

**Figure 3b.** Login

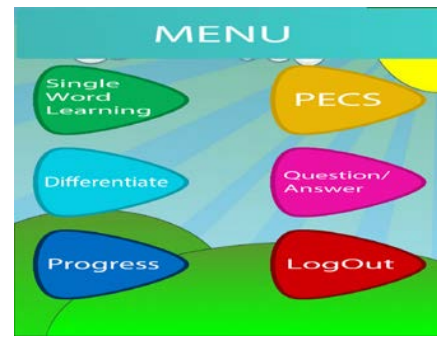

**Figure 3c.** Main menu

**Login:** This screen, as shown in Figure 3b, is for the returning users to login to the system. Upon entering valid username and password, the system authenticates and authorises a registered user to use the system and promotes to the main screen loaded with various options/functionalities.

**Main Menu:** The main menu, as shown in Figure 3c, provides with four (4) major features: Single Word learning, PECS book, Differentiate, Question & Answer. Other options include monitoring progress and logging out of the application.

**Single Word Learning:** This functionality, as shown in Figure 4a, further divides into 6 sub-categories such as Animas, Colours etc. based on relevance and grouping of the words. Each sub-categories has its own pool of pictures and audio cues accordingly e.g. if a user selects shapes, the list of different shape will be shown where user can tap every picture and listen to the audio cue associated with that picture.

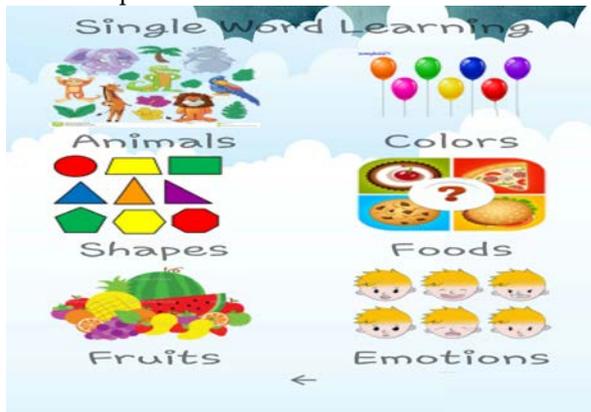

**Figure 4a.** Single word learning

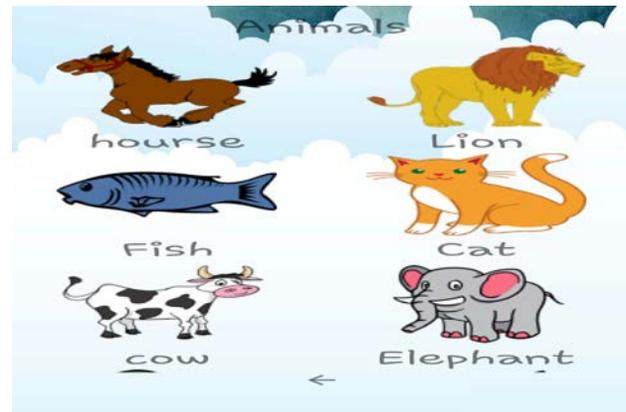

**Figure 4b.** animals

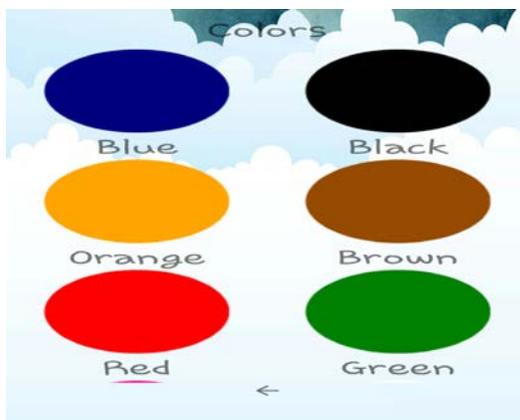

**Figure 4c.** Colours

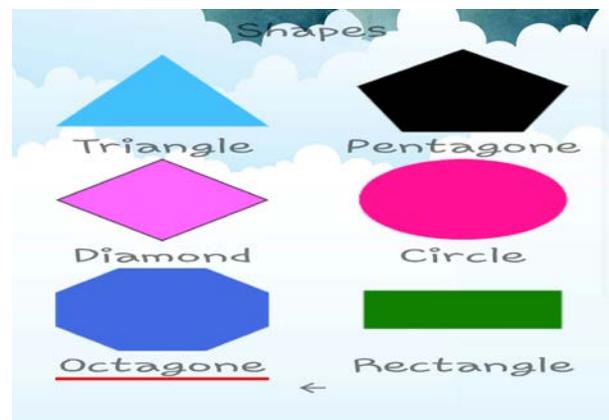

**Figure 4d.** Shapes

**Animals:** The animals screen has picture of various animals where user can tap picture and get to listen the word, as shown in Figure 4b.

**Colours:** It has pictures of different colours where the user can tap the picture and get to listen to the word, as shown in Figure 4c.

**Shapes:** This screen, as shown in Figure 4d, pictures of different shape where user can tap the picture and get to listened the word.

**Foods:** The pictures of different food, as shown in Figure 5a, where user can tap the picture and listen to the word and can learn to initiate the basic communication.

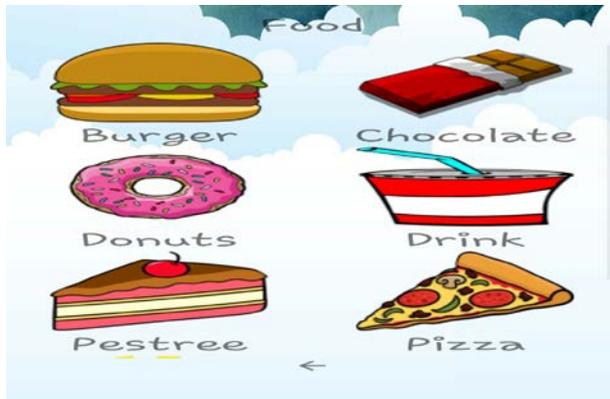

**Figure 5a.** Food

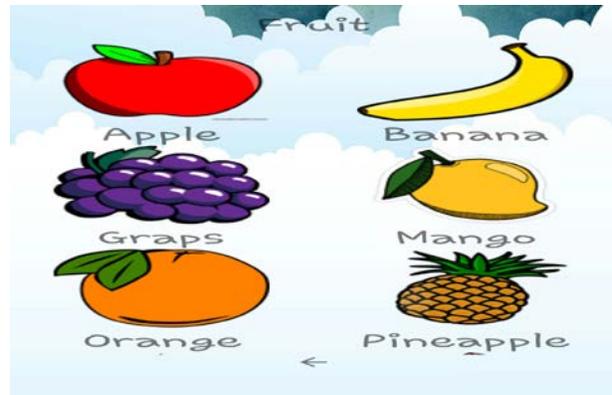

**Figure 5b.** Fruits

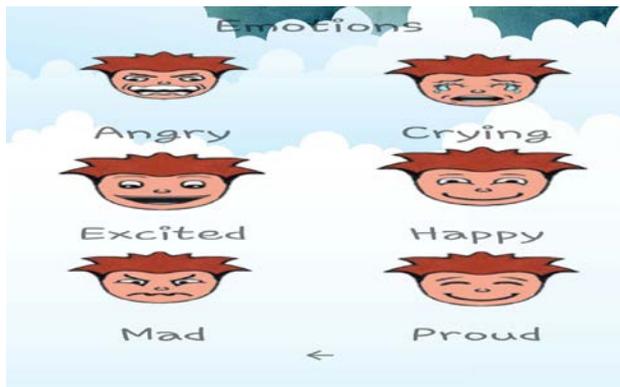

**Figure 5c.** Emotion

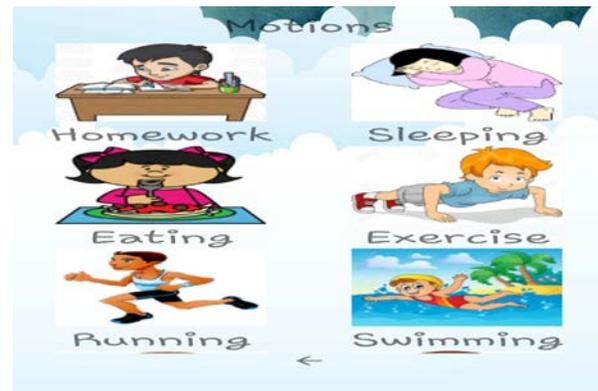

**Figure 5d.** Motion

**Fruits:** As shown in Figure 5b, the pictures of different fruits are presented so that the user can tap the picture and listen to the word and can learn to initiate the basic communication.

**Emotions:** Figure 5c demonstrates the pictures of different food where user can tap the picture and listen to the word and can learn to initiate the basic communication.

**Motions:** As shown in Figure 5d, this screen contains the pictures of different motions where user can tap the picture and listens the word and can learn to initiate the basic communication.

**Vegetables:** Figure 6a displays the Vegetables screen comprising of the pictures of different vegetables where user can tap the picture and listen to the word and can learn to initiate the basic communication.

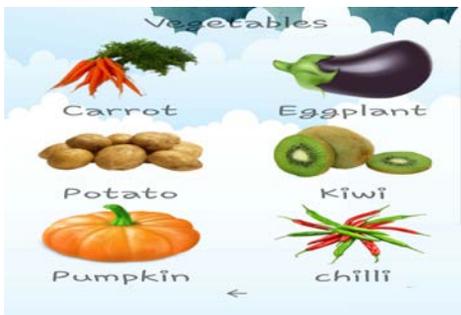

**Figure 6a.** Vegetables

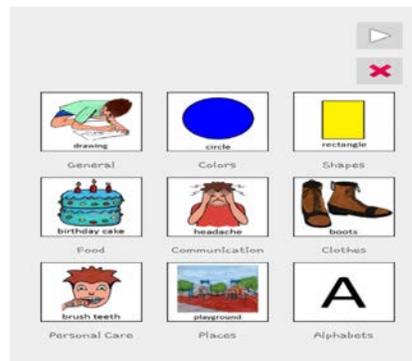

**Figure 6b.** PECS Construction book

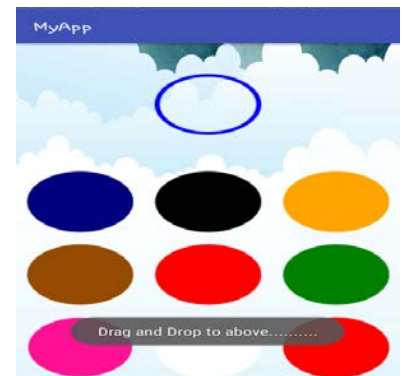

**Figure 6c.** Differentiate

**PECS Sentence Structure Book:** As demonstrated in Figure 6b, this functionality offers the autistic kids to make complete sentences followed by the picture of the item requested. For communication, children will pick multiple pictures from pictures array to create a sentence like I want food, I like to run etc. These requests also include adjectives, verbs and prepositions.

**Differentiate:** As shown in Figure 6c, the selected picture can be dragged and dropped on the matching blank space. If the selection is correct, the users will be awarded starts as points.

**Question and Answers:** As shown in Figure 7a, this screen has different questions which have three different Options to choose the right answer from. Figure 7b provides with other options while demonstrating current scores.

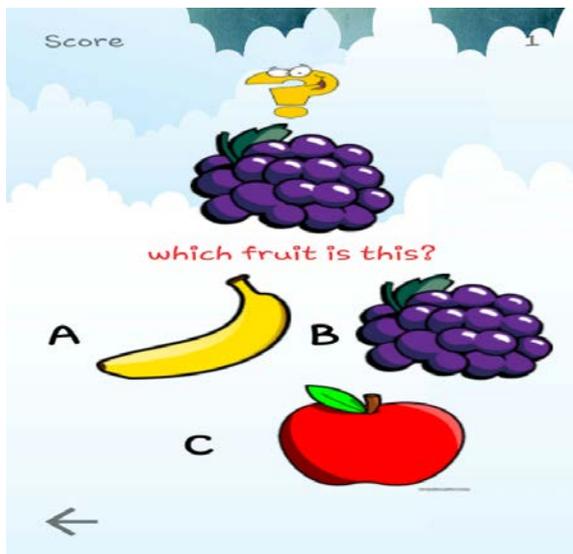

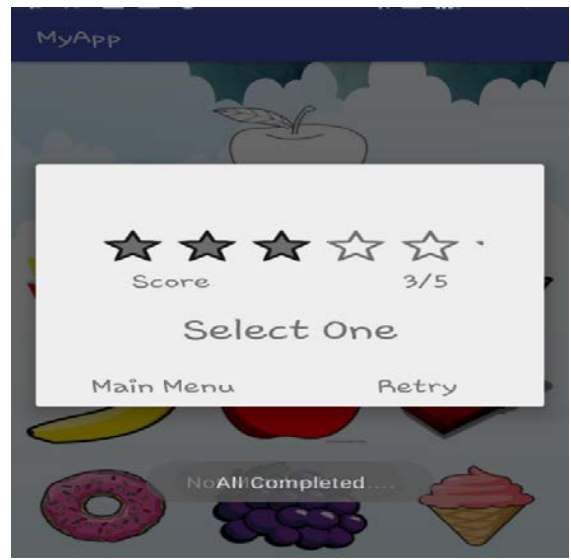

**Figure 7a.** Question and Answers                          **Figure 7b.** Scores

## 6. Experimental Evaluation of the Prototype Application (App)

While there was enthusiasm for measured the psychology of Autism kids from disordered clinic, it was imperative to conclude if studied of disorder kid should service all-purpose mechanisms. The study included analysis of the responses gained through the all-purpose questionnaires. The aim was to evaluate doctor's gratification for developing an application.

July 2017, the study conducted review of questionnaires directed at unsystematic to entitled doctors and professors from clinical psychology clinics of Campus. Doctors were distributed the questionnaires and asked to response ten assessment queries included in the questionnaire to express their observations.

Although variance in propose content, design and measurement of questionnaires, gave off an impression of being clearly better in expectations. Questions offered to understand the learning process through pictures could be essential and used for the satisfaction evaluation of speech therapists. Therefore survey and questionnaire found to be advantageous, as shown in Figure 8.

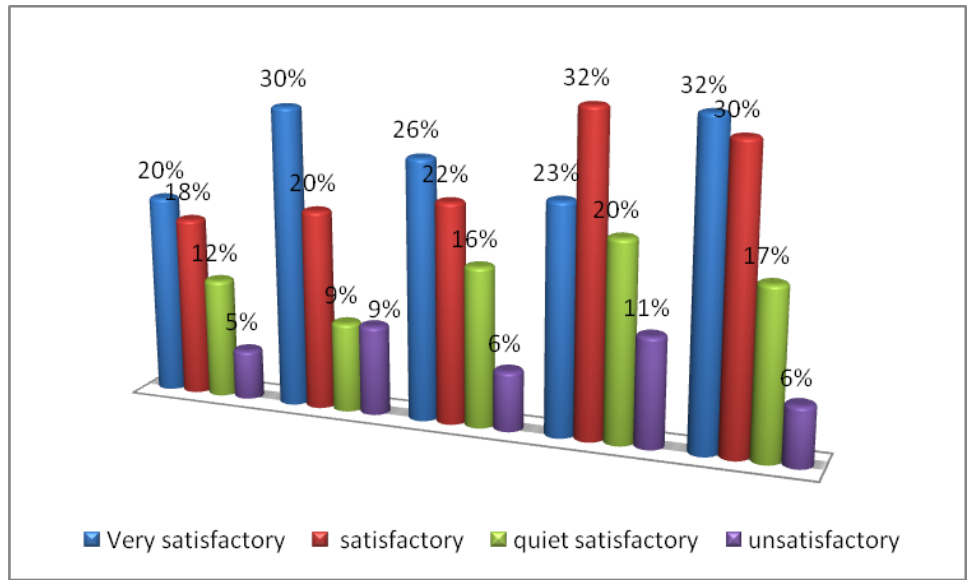

**Figure 8.** Result of the survey regarding the satisfaction

### 7. Conclusion

The research has significant contributions, in terms of both technology [24-25] and psychology, to advance active learning of the autistic kids. It is to facilitate to enable communication and instinctive among those children who suffers from Autism. Efficiency was established using autism app with kids' parent leading to possible improved household well-being. The result of the evaluation survey reveals positive outcome, satisfactory usability and improved learning capability. Future research will include furthering developing the app, moving from a prototype towards a fully-fledged cloud computing [26-27] based application.